\documentclass[preprint,prl,aps,floats]{revtex4}
\usepackage{graphicx}
\usepackage{amsmath}
\begin{document}

\title{A short note on the presence of spurious states in finite basis
approximations}
\author{R. C. Andrew and H. G. Miller \footnote {E-Mail:hmiller@maple.up.ac.za}
}
\affiliation{Department of Physics, University of Pretoria, Pretoria 0002, South
Africa}

\begin{abstract}
The genesis of spurious solutions in finite basis approximations to operators
which possess a continuum and a point spectrum is discussed  and a simple
solution for identifying these solutions is
suggested.

\end{abstract}
\maketitle

Recently Ackad and Horbatsch\cite{AH05} have presented a nice numerical method
for the solution of the Dirac equation for the hydrogenic Coulomb problem using
the Rayleigh-Ritz method\cite{KH02}.  Using a mapped Fourier grid method, a
matrix representation of the Dirac Hamiltonian is constructed in a Fourier sine
basis, which upon
diagonalization yields reasonably numerically accurate eigenvalues for a mesh
size which is not exceptionally large. Relativistic sum rules\cite{GD82} provide
a simple means of checking  whether or not the number of basis states is
adequate. As with any attempt to construct a matrix representation of an
operator which contains continuum states, spurious states can occur and must be
eliminated.  Ackad and Horbatsch\cite{AH05} have pointed out that in certain
cases they can be identified by looking at the numerical structure of the large
and small components of the corresponding eigenvector. 
Similar phenomenon occur in the mapped Fourier grid representation of the
non-relativistic
Schr{\"o}dinger problem\cite{WDM04}  in which non-physical roots are observed at
random locations.  Again the potentials considered support both bound as well as
continuum states. The wave functions of these spurious states are characterized
by their unphysical oscillations and non-vanishing amplitude in the classically
forbidden regions. The authors point out that they have found no satisfactory
mathematical explanation for the occurrence of these spurious levels.

In this note we wish to point out that the genesis of these spurious states
can
easily be understood and that there is a simple way to identify them. 
Consider an   operator, $\hat{H}$, 
which possesses  a continuum (or continua) as well as a point spectrum. The
subspace spanned  by 
its bound state eigenfunctions, $\mathcal{H_B}$,  is by itself certainly not
complete.
As the composition of this space is generally not known beforehand, a set of
basis states which is complete and spans a space, $\mathcal{F}$, is chosen to
construct a matrix representation of the operator, $\hat{H}$, to be
diagonalized.  Mathematically this corresponds to projecting the operator 
$\hat{H}$
onto the space $\mathcal{F}$. Clearly the eigenpairs obtained from
diagonalizing 
the projected operator, $\hat{H}_P$,  need not all be the same as those of the
operator $\hat{H}$.
However, because the set of basis states is complete, any state contained in
$\mathcal{H_B}$ can be expanded in terms of this set of basis states. Hence 
$\mathcal{H_B}$ may also be regarded as a subspace of $\mathcal{F}$ and the
complete
diagonalization of  $\hat{H}_P$ will yield not only the exact eigenstates of
$\hat{H}$
but additional spurious eigensolutions.  Note these spurious eigenfunctions are
eigenfunctions of $\hat{H}_P$   but not  of $\hat{H}$. Furthermore in this case
the Rayleigh-Ritz bounds discussed in the paper by Krauthauser and
Hill\cite{KH02}
apply now to the eigenstates of $\hat{H}_P$.

It is interesting to note that the same problem occurs in the Lanczos
algorithm\cite{L50} when it is applied to operators which possess a bound state
spectrum  as well as a continuum\cite{AM03}.  This is not surprising as the
Lanczos algorithm can also be considered as an application of the Rayleigh-Ritz
method\cite{P80}. In this case an orthonormalized set of Krylov basis vectors 
is used to construct iteratively a matrix represention of the operator which is 
then diagonalized.  Again  spurious states can occur for precisely the same
reasons given above.
In this case we have proposed identifying the exact bound states in the
following manner\cite{AM03}. After  
each iteration,  for each of the converging eigenpairs ($e_{l \lambda}$,$|e_{l 
\lambda}\rangle$), $\Delta_{l \lambda}=|e_{l \lambda}^2-<e_{l 
\lambda}|\hat{H}^2|e_{l \lambda}>|$ (where $l$ is the iteration number) is 
calculated and a determination is made  as to whether  $\Delta$ is converging  
toward zero or not.  For the exact bound states of $\hat{H}$,  $\Delta$ must be 
identically zero while the  other eigenstates states of the projected operator 
should converge  to some non-zero positive   value. This method has been
successfully implemented to identify spurious states in
non-relativistic\cite{AM03}
as well as relativistic\cite{AMY07} eigenvalue problems. A similar procedure can
be implemented in any Rayleigh-Ritz application.  One simply must check to see
whether the eigensolutions from the diagonalization  of $\hat{H}_P$  are also
eigensolutions of $\hat{H}^2$.

\end{document}